\documentclass[aps,prd,twocolumn,showpacs,amsmath,amssymb]{revtex4-1}
\usepackage{amsmath}
\usepackage{graphicx}
\usepackage{subfigure}
\usepackage{epstopdf}
\usepackage{color}
\usepackage{xcolor}
\usepackage{multirow}
\usepackage{setspace}
\usepackage{overpic}
\usepackage{amssymb}
\usepackage[bookmarksnumbered, pdfstartview=FitH,colorlinks,urlcolor=blue, citecolor=blue,linkcolor=blue] {hyperref}
\usepackage[mathlines]{lineno}
\usepackage{bm}
\usepackage{rotating}
\usepackage{siunitx}
\usepackage[utf8]{inputenc}
\begin{document}
	
	%\linenumbers
	
	\title{\bf \boldmath{
			Search for the lepton number violating decay $\eta\to \pi^+\pi^+e^-e^- + c.c.$ via $J/\psi\to\phi\eta$}
	}

\author{
M.~Ablikim$^{1}$, M.~N.~Achasov$^{4,c}$, P.~Adlarson$^{77}$, X.~C.~Ai$^{82}$, R.~Aliberti$^{36}$, A.~Amoroso$^{76A,76C}$, Q.~An$^{73,59,a}$, Y.~Bai$^{58}$, O.~Bakina$^{37}$, Y.~Ban$^{47,h}$
% ...
\\
\vspace{0.2cm}
(BESIII Collaboration)\\
\vspace{0.2cm} {\it
$^{1}$ Institute of High Energy Physics, Beijing 100049, People's Republic of China\\
% ...
$^{4}$ Budker Institute of Nuclear Physics SB RAS (BINP), Novosibirsk 630090, Russia\\
% ...
$^{36}$ Johannes Gutenberg University of Mainz, Johann-Joachim-Becher-Weg 45, D-55099 Mainz, Germany\\
$^{37}$ Joint Institute for Nuclear Research, 141980 Dubna, Moscow region, Russia\\
% ...
$^{47}$ Peking University, Beijing 100871, People's Republic of China\\
% ...
$^{58}$ Southeast University, Nanjing 211100, People's Republic of China\\
$^{59}$ State Key Laboratory of Particle Detection and Electronics, Beijing 100049, Hefei 230026, People's Republic of China\\
% ...
$^{73}$ University of Science and Technology of China, Hefei 230026, People's Republic of China\\
% ...
$^{76}$ University of Turin and INFN, (A)University of Turin, I-10125, Turin, Italy; (B)University of Eastern Piedmont, I-15121, Alessandria, Italy; (C)INFN, I-10125, Turin, Italy\\
$^{77}$ Uppsala University, Box 516, SE-75120 Uppsala, Sweden\\
% ...
$^{82}$ Zhengzhou University, Zhengzhou 450001, People's Republic of China\\
\vspace{0.2cm}
$^{a}$ Deceased\\
% ...
$^{c}$ Also at the Novosibirsk State University, Novosibirsk, 630090, Russia\\
% ...
$^{h}$ Also at State Key Laboratory of Nuclear Physics and Technology, Peking University, Beijing 100871, People's Republic of China\\
% ...
}
}

	\begin{abstract} %OK
		Based on a sample of $ (10.087\pm 0.044)\times 10^{9} J/\psi$ events collected by the BESIII detector at the BEPCII collider, we perform the first search for the lepton number violating decay $\eta \to \pi^+\pi^+ e^-e^- + \text{c.c.}$  No signal is found, and an upper limit on the branching fraction of $\eta \to \pi^+\pi^+ e^-e^- + c.c.$ is set to be $4.6 \times 10^{-6}$ at the 90\% confidence level.
	\end{abstract}
	
	\maketitle

	\section{Introduction} %OK
	Searches for rare decays have been instrumental in the development and validation of the Standard Model (SM). A prime example is the investigation of long-lived neutral kaons decaying into two pions, which ultimately led to the groundbreaking discovery of $CP$ violation~\cite{CPviolation}.
	Today, rare or forbidden decays continue to challenge experimental precision and theoretical rigor,  highlighting the need for more extensive and in-depth studies to further our understanding of phenomena beyond the SM.
	
	A critical goal in the exploration of rare and forbidden decays is the pursuit of lepton number violating (LNV) processes. Within the SM, the total lepton number~($L$) is conserved. However, the phenomenon of neutrino oscillation suggests that neutrinos possess non-zero mass. According to the SU(2)$\otimes$U(1) theory concerning left-handed neutrino fields~\cite{ref:spontaneousLNV}, massive neutrinos could induce spontaneous LNV processes. 
	An alternative route to LNV involves the introduction of new heavy particles beyond the scope of the SM, facilitated through an effective Lagrangian of dimensionality five, which permits LNV processes~\cite{Weinberg_unconserve}. One such candidate is the Majorana neutrino~\cite{Majorana_ori}, whose antiparticle is the neutrino itself, leading to LNV processes characterized by a lepton-number change of two~($|\Delta L|=2$). Consequently, the detection of an LNV process with $|\Delta L| = 2$ would provide compelling evidence for the existence of Majorana neutrinos~\cite{theo_0vbb1,theo_0vbb2}. The seesaw mechanism posits that the small masses of SM neutrinos could originate from much larger Majorana masses~\cite{seesaw1, seesaw2, seesaw3, seesaw4, seesaw5, ref:nvmass1}. In this case, the search for LNV processes becomes crucial for elucidating the origin of neutrino mass and deepening our understanding of these fundamental particles.
	
	Under the current constraints of the SM, lepton number and baryon number~($B$) follow $|\Delta(B-L)| = 0$, which means LNV processes also imply baryon number violation~(BNV). If BNV processes did not occur during the early Universe, the quantities of matter and antimatter should have remained constant. Consequently, processes with both LNV and BNV could offer a compelling explanation for the observed baryon-antibaryon asymmetry in the Universe. 
	
	Over the past few years, numerous collider experiments, including LHCb~\cite{LNV_experiment1}, CMS~\cite{LNV_experiment2}, BaBar~\cite{BabarLNV2021}, ATLAS~\cite{LNV_experiment4}, CLEO~\cite{CLEOLNV2005}, FOCUS~\cite{FOCUSLNV2003}, and BESIII~\cite{LNV_experiment7}, have performed extensive searches for LNV processes. Nuclear experiments such as CDEX-1B\cite{CDEX2023} and KamLAND-Zen\cite{KamLAND-Zen:2024eml} are also searching for LNV processes. Despite these diligent efforts, no significant evidence of an LNV signal has yet been detected, leaving the quest for LNV processes and the physics beyond the SM an open and active area of research.
	In the context of the Majorana neutrino, these processes are anticipated to be extremely suppressed if the LNV propagator has a heavy mass. Nonetheless, the detection of any indication within the current data samples would unequivocally point towards the presence of new physics, potentially shedding light on the nature of neutrinos and underlying mechanisms beyond the SM.

	In this analysis, we perform the first search for a LNV decay of the $\eta$ meson.  In particular, we search for the SM forbidden decay $\eta\to \pi^+\pi^+e^-e^-$ (charged conjugate modes are implied throughout this paper). This analysis is performed using a data  sample of $10.087\times10^{9}$ $J/\psi$ events~\cite{BESIII:Jpsiyields} collected with the BESIII detector operating at the BEPCII storage ring.

	\section{BESIII detector and Monte Carlo simulation} %OK
	The BESIII detector~\cite{BESIII}  records symmetric $e^+e^-$ collisions provided by the BEPCII collider~\cite{Yu:IPAC2016-TUYA01} in the center-of-mass energy range from 1.84 to 4.95 GeV, with a peak luminosity of $1.1\times10^{33}{\rm cm^{-2}s^{-1}}$ achieved at $\sqrt{s} = 3.773$ GeV. BESIII has collected large data samples in this energy region~\cite{BESIIIdata1,BESIIIdata2,BESIIIdata3}. The cylindrical core of the BESIII detector covers 93\% of the full solid angle and consists of a helium-based multilayer drift chamber (MDC), a plastic scintillator time-of-flight system (TOF), and a CsI (Tl) electromagnetic calorimeter (EMC), which are all enclosed in a superconducting solenoidal magnet providing a 1.0~T magnetic field. The magnetic field intensity was 0.9~T during the 2012 data-taking period (below nominal specifications), affecting 11\% of the total $J/\psi$ data samples.
	
	The solenoid is supported by an octagonal flux-return yoke with resistive plate counter muon identifier modules interleaved with steel. The charged-particle momentum resolution at $1~{\rm GeV}/c$ is $0.5\%$, and the resolution of the specific ionization energy loss (d$E$/d$x$) is $6\%$ for the electrons from Bhabha scattering. The EMC measures photon energies with a resolution of $2.5\%$ ($5\%$) at $1$~GeV in the barrel (end cap) region. The time resolution of the TOF barrel part is 68~ps, while that of the end cap part is 110~ps. The end cap TOF system was upgraded in 2015 using multi-gap resistive plate chamber technology, providing a time resolution of 60 ps~\cite{60ps1,60ps2}. Approximately 87\% of the data used in this analysis was collected after this upgrade. More details about the design and performance of the BESIII detector are given in Ref.~\cite{BESIII}.
	
	Simulated samples, produced with the {\sc geant4}-based~\cite{geant4} Monte Carlo (MC) package including the geometric description of the BESIII detector and the
	detector response, are used to determine the detection efficiency and to estimate the backgrounds. The simulation includes the beam-energy spread and initial-state radiation in the $e^+e^-$ annihilations modeled with the generator {\sc kkmc}~\cite{kkmc1}. The inclusive MC samples consist of the production of the $J/\psi$ and the continuum processes. The known decay modes are modeled with {\sc evtgen}~\cite{evtgen} using branching fractions taken from the Particle Data Group (PDG)~\cite{PDG2022}, and the remaining unknown decays from the charmonium states are modeled with {\sc lundcharm}~\cite{lundcharm,lundcharm2}. The final-state radiation from charged final-state particles is incorporated with the {\sc photos} package~\cite{photos}.
	
	\section{METHOD} 
	In this analysis, we search for the decay $\eta\to\pi^+\pi^+e^-e^-$ via $J/\psi\to\phi\eta$ with $\phi\to K^+K^-$.  To avoid possible bias, a blind analysis technique is performed where the full data set is analyzed only after the analysis procedure is fixed and validated with MC simulation. In order to avoid the large uncertainty from $\mathcal{B}(J/\psi\to\phi\eta)$~\cite{PDG2022}, which is about 11\%, we measure the branching fraction of the signal decay $\eta\to\pi^+\pi^+e^-e^-$ relative to that of the reference channel $\eta\to\gamma\gamma$. Here the uncertainty of the input $\mathcal{B}(\eta\to \gamma\gamma)$ is only $0.5\%$~\cite{PDG2022}.

	The branching fractions of $\eta\to\pi^+\pi^+e^-e^-$ and $\eta\to\gamma\gamma$ can be expressed as
	\begin{linenomath*}
		\begin{equation}
			\label{eq:1}
			\mathcal{B}(\eta\to\pi^+\pi^+e^-e^-)=
			\frac{N^{\rm net}_{\pi^+\pi^+e^-e^-}/\epsilon_{\pi^+\pi^+e^-e^-}}{N^{\rm tot}\cdot\mathcal{B}_{\rm offset}}
		\end{equation}
	\end{linenomath*}
	and
	\begin{linenomath*}
		\begin{equation}
			\label{eq:2}
			\mathcal{B}(\eta\to\gamma\gamma)=\frac{N^{\rm net}_{\gamma\gamma}/\epsilon_{\gamma\gamma}}{N^{\rm tot}\cdot\mathcal{B}_{\rm offset}}
		\end{equation}
	\end{linenomath*}
	respectively, where $\mathcal{B}_{\rm offset}=\mathcal{B}(J/\psi\to\phi\eta)\cdot\mathcal{B}(\phi\to K^+K^-)$,  $N^{\rm net}_{\pi^+\pi^+e^-e^-}$ and $N^{\rm net}_{\gamma\gamma}$ represent the net yields for the two $\eta$ decay channels, $N^{\rm tot}$ denotes the total number of $J/\psi$ events in data, and $\epsilon_{\pi^+\pi^+e^-e^-}$ and $\epsilon_{\gamma\gamma}$ are the detection efficiencies for the signal and reference channels, respectively. The terms $\mathcal{B}(J/\psi \to \phi\eta)$ and $\mathcal{B}(\phi \to K^+K^-)$ represent the individual branching fractions. Utilizing these two equations, the branching fraction of $\eta \to \pi^+\pi^+e^-e^-$ can be reformulated as%
	\begin{linenomath*}
		\begin{equation}
			\label{eq:reference}
			\mathcal{B}(\eta\to\pi^+\pi^+e^-e^-) = \mathcal{B}(\eta\to \gamma\gamma)
			\cdot\frac{N^{\rm net}_{\pi^+\pi^+e^-e^-}\cdot \epsilon_{\gamma\gamma}}{\epsilon_{\pi^+\pi^+e^-e^-}\cdot N^{\rm net}_{\gamma\gamma}}.
		\end{equation}
	\end{linenomath*}
	
	%The uncertainty of the input $\mathcal{B}(\eta\to \gamma\gamma)$ is only $0.5\%$~\cite{PDG2022}, significantly reducing the total systematic uncertainty. 
	
	\section{Analysis of $\eta\to\pi^+\pi^+e^-e^-$} \label{sec::LNV} %OK
	
	In each event, at least six charged tracks are required based on the hypothesis of $e^+e^- \to K^+K^-\pi^+\pi^+e^-e^-$. All charged tracks detected in the MDC must meet a polar angle criterion of $\vert\!\cos\theta\vert < 0.93$, which corresponds to the maximum detection range of the MDC. The tracks must originate from the interaction point, with the distance along the $z$ axis, $|V_z|$, being less than 10~cm, and in the transverse plane, the distance, $V_{xy}$, must be less than 1~cm.
	For the particle identification (PID) of charged tracks, we make use of the d$E$/d$x$ measured in the MDC, the time of flight, and the energy deposition and cluster shape in the EMC. These parameters are utilized to compute the confidence levels (CLs) for the electron, kaon, and pion hypotheses, denoted as $\mathrm{CL}_e$, $\mathrm{CL}_K$ and $\mathrm{CL}_\pi$, respectively. 
	Charged tracks with $\mathrm{CL}_e > 0.001$ and $\frac{\mathrm{CL}_e}{\mathrm{CL}_e + \mathrm{CL}_K + \mathrm{CL}_\pi} > 0.8$ {{are assigned as electron candidates, those with  $\mathrm{CL}_{K} > 0.001$ and $\mathrm{CL}_{K} > \mathrm{CL}_{\pi}$ are assigned as kaon candidates, and those with $\mathrm{CL}_\pi > \mathrm{CL}_K$ are assigned as pion candidates. }}
	
	To suppress background events and improve mass resolution, a kinematic fit is performed by constraining the total four-momentum (4C) to the initial $e^+e^-$ four-momentum. All six charged tracks are included in the kinematic fit. The combination with the smallest  $\chi^2_{\rm 4C}$ is selected for further analysis. By utilizing the punzi method~\cite{punzi}, the optimal $\chi^2_{\rm 4C}$ for the signal channel is determined to be less than 20.
	
	The two-dimensional~(2D) signal region is defined as $M_{\pi^+\pi^+e^-e^-}\in [0.530, 0.564]~{\rm GeV}/c^{2}$ and $M_{K^+K^-}\in [1.008, 1.031]~{\rm GeV}/c^{2}$, %in the $M_{\pi^+\pi^+e^-e^-}$ and $M_{K^+K^-}$ distributions, respectively, 
	based on a study of signal MC events.  In the signal MC study, a double Gaussian function is used to represent the signal and second-order Chebyshev polynomial functions are used to represent the background. The signal region corresponds to $\pm$3 times the mass resolution centered around the known $\eta$ and $\phi$ masses~\cite{PDG2022}, for $M_{\pi^+\pi^+e^-e^-}$ and $M_{K^+K^-}$, respectively. The detection efficiency is calculated to be 11.20\% using the signal MC simulation. In these simulations, the $J/\psi$ decay is modeled using the helicity amplitude generator HELAMP, while the $\phi$ and $\eta$ decays are modeled using the VSS (generator of a vector particle decaying into a pair of scalars) and phase-space (PHSP) generators~\cite{evtgen,evtgen2}.
	
	\begin{figure}[htbp]
		\centering
		\includegraphics[width=0.45\textwidth]{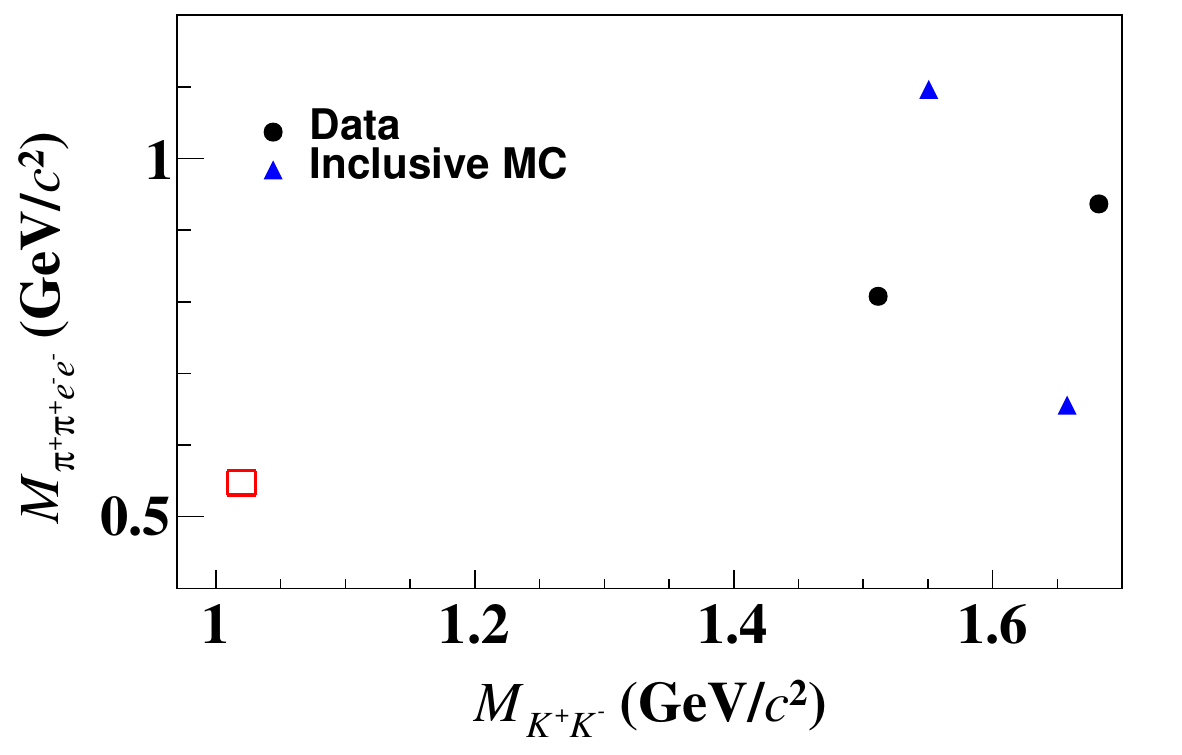}
		\caption{The distribution of $M_{K^+K^-}$ versus $M_{\pi^+\pi^+e^-e^-}$, where the black solid points are data, the blue triangles represent the inclusive MC sample and the red rectangle indicates the signal region. }
		\label{fig:data-inc}
	\end{figure}
	
	Utilizing the event type analysis tool, TopoAna~\cite{topoanaAlg}, the backgrounds originating from $J/\psi$ decays are examined using an inclusive MC sample. It is found that only two events remain, and both of them are located well outside the signal region.

	The 2D distribution of $M_{K^+K^-}$ versus $M_{\pi^+\pi^+e^-e^-}$ for the background events from the inclusive MC sample is shown in Fig.~\ref{fig:data-inc}. The red rectangle in the figure shows the signal region, and it is found that no background events are  present around the signal region.

	\section{Analysis of $\eta\to \gamma\gamma$} %OK
	
	To eliminate the uncertainty associated with the decay $J/\psi \to \phi\eta$, we reconstruct the reference channel using $\eta \to \gamma\gamma$ and $\phi \to K^+K^-$. In each event, we require two good charged tracks and at least two photon candidates.
	The analysis strategy for selecting the good charged and neutral tracks, together with identifying the charged kaons, are identical to those described in Sec.~\ref{sec::LNV}. 

	Photon candidates are identified from isolated clusters in the EMC. To mitigate the impact of electronic noise and beam-related background, clusters must start within 700~ns of the event start time and must be positioned outside a cone angle of $20^\circ$ around the nearest extrapolated good charged track. The minimum energy threshold for each EMC cluster is set to 25 MeV for the barrel region ($\vert\!\cos\theta\vert < 0.80$) and 50 MeV for the end cap regions ($0.86 < \vert\!\cos\theta\vert < 0.92$). 
	
	The 4C kinematic fit is performed based on the hypothesis of $e^+e^- \to K^+K^-\gamma\gamma$, and is optimized with a figure of merit~(FOM) of the form $S/\sqrt{S+B}$, where $S$ and $B$ are the signal and background yields, respectively. The $\chi^2_{\rm 4C}$ for the reference channel is required to be less than 40.

\begin{figure}[htbp]
		\centering
		\includegraphics[width=0.4\textwidth]{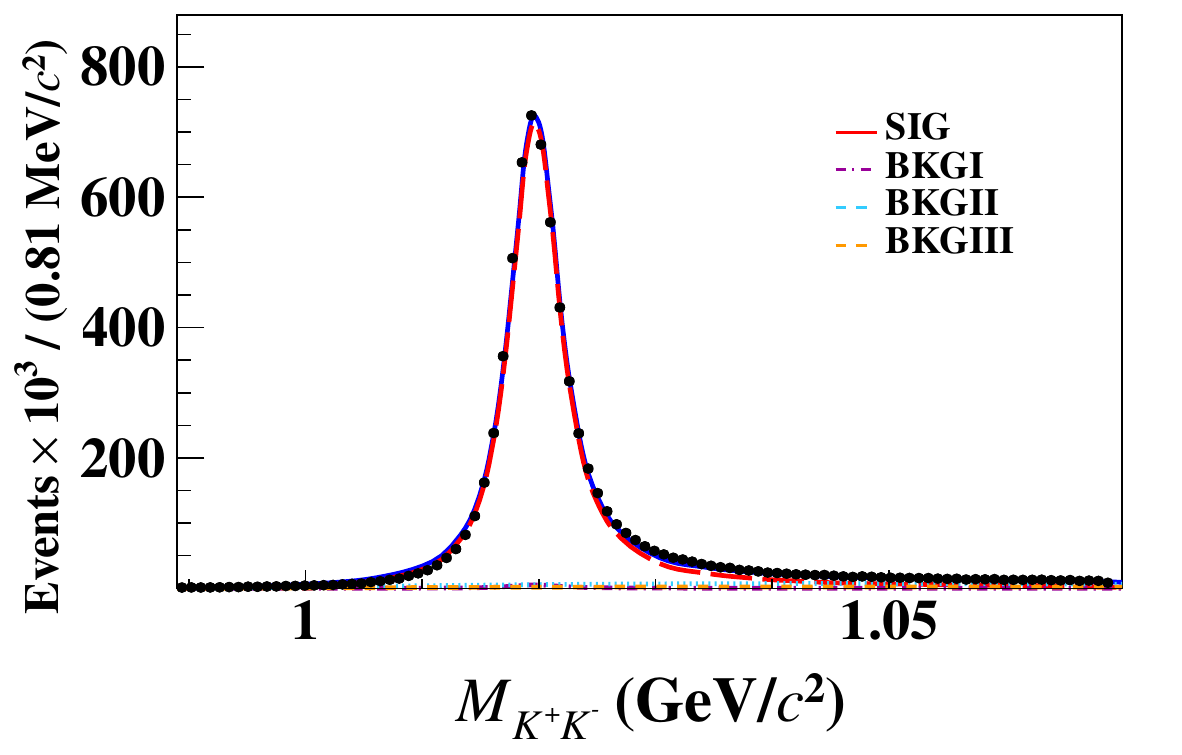}
		\includegraphics[width=0.4\textwidth]{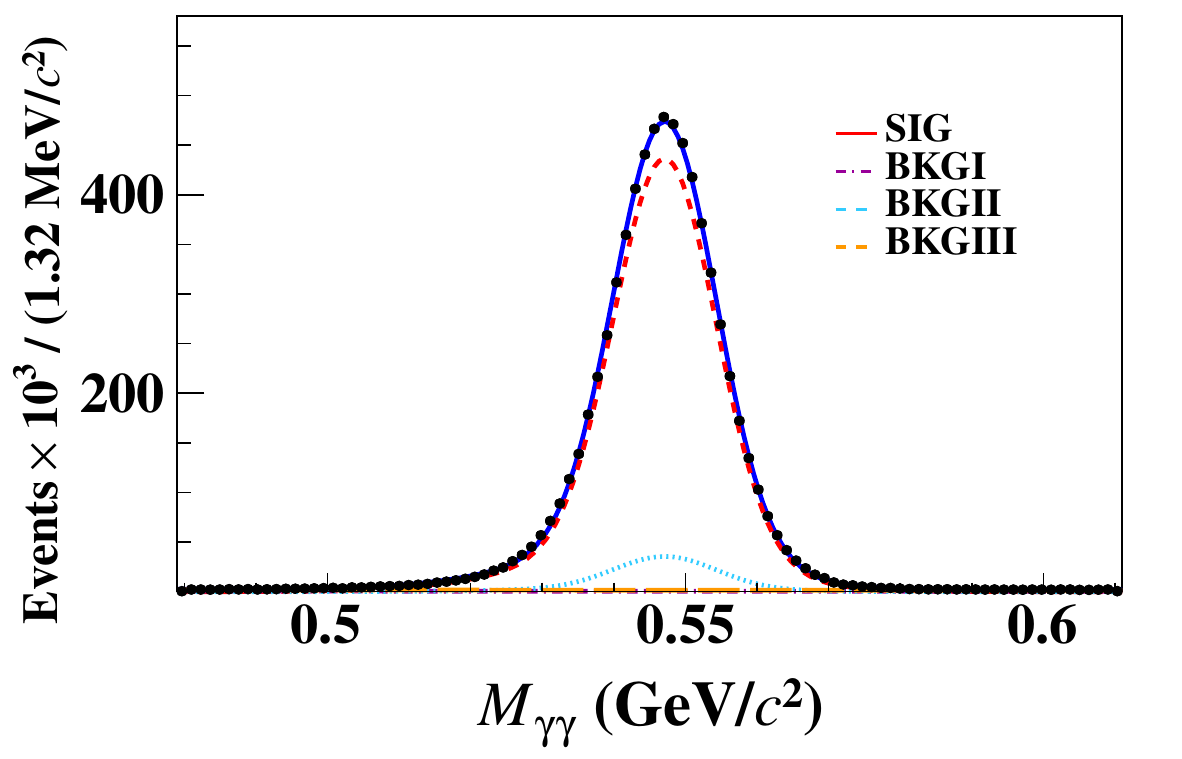}
		\caption{Projections of the 2D fit to the distributions of $M_{K^+K^-}$ and $M_{\gamma\gamma}$ in data. The black dots are data, the blue line is the total fit result, the red line is the SIG component, the purple dashed line is the BKGI component, the blue dashed line is the BKGII component, and the orange dashed line is the BKGIII component.}
		\label{fig:2D-fit}
	\end{figure}

	To determine the signal yield for the reference channel $J/\psi \to \phi\eta, \phi \to K^+ K^-, \eta \to \gamma\gamma$, we perform a 2D fit to the distribution of $M_{K^+K^-}$ versus $M_{\gamma\gamma}$. The fit region for $M_{K^+K^-}$ is defined as $[0.989, 1.069]~{\rm GeV}/c^2$, and for $M_{\gamma\gamma}$ as $[0.480, 0.610]~{\rm GeV}/c^2$. These fit ranges are set according to $\mu \pm 9\sigma$, where $\mu$ and $\sigma$ are the mean and standard deviation of the Gaussian fit to the $M_{\gamma\gamma}$ or $M_{K^+ K^-}$ distribution.
	In the 2D fit, the $\phi$ signal is modeled using a truth-matched MC shape (shape1), the $\eta$ signal is modeled using a double Gaussian function (shape2), the background for $M_{K^+ K^-}$ is described by an inverse ARGUS function~\cite{argus} (shape3), and the background projection for $M_{\gamma\gamma}$ is described by a third-order polynomial function (shape4). Consequently, the total signal shape is represented by $\mathrm{shape1} \otimes \mathrm{shape2}$ (SIG), potential backgrounds like $\phi\gamma\gamma$ are represented by $\mathrm{shape1} \otimes \mathrm{shape4}$ (BKGI), backgrounds like $\eta K^+K^-$ are represented by $\mathrm{shape2} \otimes \mathrm{shape3}$ (BKGII), and the flat background is described by $\mathrm{shape3} \otimes \mathrm{shape4}$ (BKGIII).
	Based on the 2D fit shown in Fig.~\ref{fig:2D-fit}, the number of $\eta \to \gamma\gamma$ candidate events is determined to be $N^{\rm net}_{\gamma\gamma}=(647.5 \pm 0.9)\times 10^{3}$.
	
	The detection efficiency is estimated with simulated signal events, where the decays $J/\psi \to \phi\eta$, $\phi \to K^+K^-$, and $\eta \to \gamma\gamma$ are modeled using HELAMP, the VSS model, and the PHSP generator, respectively.  The detection efficiency is calculated to be $\epsilon_{\gamma\gamma}=42.15\%$. The corresponding branching fraction of $J/\psi\to \phi\eta$ is $\mathcal{B}(J/\psi\to \phi\eta)=(7.7\pm0.4)\times10^{-4}$. This result does not consider the interference between $J/\psi$ decay and QED continuum process and is consistent with the previous measurements within errors\cite{PDG2022, BESIII:JpsiPhietaphase}.

	\section{Systematic uncertainty} %OK
	
	The measured branching fraction is subject to various sources of systematic uncertainty, including tracking, PID, 4C kinematic fit, signal window, fit procedure, MC modeling, MC statistics, determination of $N^{\rm net}_{\gamma\gamma}$, and the branching fraction $\mathcal{B}(\eta\to\gamma\gamma)$. A summary of all systematic uncertainties is provided in Table~\ref{tab:uncertainty}. The total uncertainty is calculated to be 7.5\% by summing the individual components in quadrature.

	\begin{table}[htbp]
		\centering
		\caption{Relative systematic uncertainties in the branching fraction measurement.}
		\label{tab:uncertainty}
		\begin{tabular}{lc}
			\hline\hline
			Source                             & Uncertainty~(\%) \\\hline
			MDC tracking                       & 4.0\\  
			PID                                & 4.0\\
			Photon detection                   & 2.0\\
			4C kinematic fit   & {3.8}\\
			2D fit                             & 2.0\\
			Signal window                      & 0.2\\
			MC modeling                & 0.6\\
			MC statistics                      & 0.9\\
			$\mathcal{B}(\eta\to\gamma\gamma)$ & 0.5\\\hline
			Total                              & 7.5\\\hline
			\hline
		\end{tabular}
	\end{table}

	\begin{itemize}
		
		\item The uncertainties associated with the tracking and PID efficiencies for the $K^{\pm}$, $\pi^{\pm}$, and $e^{\pm}$ are assessed using control samples. For electrons, the process $e^+e^- \to \gamma e^+e^-$ is used. The corresponding uncertainties for kaons and pions are investigated using $\psi(3686) \to \pi^+\pi^-J/\psi$ and $J/\psi \to K(892)^{*0}K^0_S \to K^0_SK^+\pi^- \to K^+\pi^-\pi^+\pi^-$ events. The discrepancies in tracking/PID efficiencies between data and MC simulation for $K^\pm$ ($\pi^\pm$), referred to as data-MC differences, are assigned as the uncertainties~\cite{trackuncertainty}. Subsequently, to be conservative, the systematic uncertainty for tracking/PID is assigned as 1\% for each kaon, pion, and electron. Since both the signal and reference channels involve the decay $\phi \to K^+K^-$, the tracking and PID uncertainties associated with $K^+$ and $K^-$ cancel. After adding the systematic uncertainties of each track, the total systematic uncertainties for MDC tracking or PID is assigned to be 4\%.
		
		\item The photon detection uncertainty is studied using a control sample of $J/\psi\to\rho^0\pi^0$ . The data-MC difference of photon selection efficiencies is assigned as 1.0\% per photon. The systematic uncertainty due to photon reconstruction is assigned to be 2.0\% for two photons.
		
		\item The systematic uncertainty due to the 4C kinematic fit for $J/\psi\to\phi\eta\to K^+K^-\eta(\eta\to\gamma\gamma)$ and $\chi^2_{\rm 4C}<40$ is studied using a control sample of $J/\psi\to K^+K^-\pi^0(\gamma\gamma)$. The corresponding uncertainty is estimated to be 2.4\% by comparing the difference of efficiencies between data and MC simulation. Similarly, the systematic uncertainty due to the 4C kinematic fit and $\chi^2_{\rm 4C}<20$ for $J/\psi\to\phi\eta\to K^+K^-\eta(\eta\to\pi^+\pi^+e^-e^-)$ is studied using a control sample of $J/\psi\to K^+K^-\pi^+\pi^-\pi^+\pi^-$. The corresponding uncertainty is assigned to be 2.9\%. Finally, the systematic uncertainty due to the 4C kinematic fit is assigned to be 3.8\%. 
		
		\item The uncertainty in the 2D fit to $J/\psi\to\phi\eta\to K^+K^-\eta(\gamma\gamma)$ is estimated by changing the signal and background parameters by $\pm 1\sigma$. The parameters include the width of the double Gaussian function and the end-point of the ARGUS function.  The maximum difference is assigned as the uncertainty, which is 2.0\%.
		
		\item To determine the uncertainty due to the choice of the $\eta$ signal window, we use different signal windows, including $\pm 3.2\sigma$, $\pm 3.0\sigma$, $\pm 2.8\sigma$, etc. The standard deviation on the detection efficiency of 0.2\% is taken as the uncertainty.
		
		\item To estimate the uncertainty due to the MC modeling, we change the signal MC generator used for the decay $\phi \to K^+K^-$ from the VSS model to the PHSP model. The change in the signal efficiency, which is 0.6\%, is taken as the MC model uncertainty.

		\item The uncertainty due to the MC statistics, 0.9\%, is given by $\Delta\epsilon/\epsilon=\sqrt{\frac{\epsilon(1-\epsilon)}{N^{\rm MC}_{\rm gen}}}/\epsilon$, where $\epsilon$ is the corresponding detection efficiency, and $N^{\rm MC}_{\rm gen}$ is the total number of generated signal MC events. 
		
		\item The uncertainty of the quoted branching fraction $\mathcal{B}(\eta\to \gamma\gamma)$ is 0.5\%. 
		
	\end{itemize}

	\section{Result} 
	
	Since there is no signal or background event observed in the signal region, the upper limit of the signal yield $N^{\rm up}_{\pi^+\pi^+e^-e^-}$ is estimated to be 17.81 at the 90\% CL. The upper limit is set using a frequentist method~\cite{TROLKE} with an unbounded profile likelihood treatment of systematic uncertainties, where the background fluctuation is assumed to follow a Poisson distribution, the detection efficiency ($\epsilon=11.2\%$) is assumed to follow a Gaussian distribution, and the systematic uncertainty ($\Delta_{\rm sys}=7.5\%$) is considered as the standard deviation of the efficiency.
	
	The upper limit of the branching fraction of $\eta\to\pi^+\pi^+e^-e^-$ is determined to be
	\begin{linenomath*}
		\begin{equation}
			\mathcal{B}(\eta\to\pi^+\pi^+e^-e^-)<\mathcal{B}(\eta\to\gamma\gamma)\times\frac{N^{\rm up}_{\pi^+\pi^+e^-e^-}}{N^{\rm net}_{\gamma\gamma}/\epsilon_{\gamma\gamma}},
		\end{equation}
	\end{linenomath*}
	where $\epsilon_{\gamma\gamma}=42.15\%$, { $N^{\rm net}_{\gamma\gamma}=(647.5\pm0.9)\times 10^3$}, and $\mathcal{B}(\eta\to\gamma\gamma)=(39.36\pm 0.18)\%$. Thus, the resulting upper limit of the branching fraction at
	the 90\% SL is set to be
	\begin{linenomath*}
		\begin{equation}
			\mathcal{B}(\eta\to\pi^+\pi^+e^-e^-)<4.6\times10^{-6}.
		\end{equation}
	\end{linenomath*}

\section{Summary} 
In summary, by analyzing a sample of $(10.087 \pm 0.044) \times 10^{9} J/\psi$ events collected with the BESIII detector at the BEPCII collider, we perform the first search for the LNV decay $\eta \to \pi^+\pi^+e^-e^-$ via $J/\psi \to \phi\eta$. No signal is observed in the data, and the upper limit of its branching fraction is set to be $4.6 \times 10^{-6}$ at the 90\% CL. This is the first experimental constraint on LNV signals in $\eta$ decays and is complementary to the constraints placed by dedicated experiments, thereby enhancing our understanding of the LNV process in light hadrons from an experimental perspective.

\section{Acknowledgement}
The BESIII Collaboration thanks the staff of BEPCII (https://cstr.cn/31109.02.BEPC) and the IHEP computing center for their strong support. This work is supported in part by National Key R\&D Program of China under Contracts Nos. 2023YFA1606000, 2023YFA1606704; National Natural Science Foundation of China (NSFC) under Contracts Nos. 12035009, 11635010, 11935015, 11935016, 11935018, 12025502, 12035013, 12061131003, 12192260, 12192261, 12192262, 12192263, 12192264, 12192265, 12221005, 12225509, 12235017, 12361141819; the Chinese Academy of Sciences (CAS) Large-Scale Scientific Facility Program; CAS under Contract No. YSBR-101; 100 Talents Program of CAS; The Institute of Nuclear and Particle Physics (INPAC) and Shanghai Key Laboratory for Particle Physics and Cosmology; German Research Foundation DFG under Contract No. FOR5327; Istituto Nazionale di Fisica Nucleare, Italy; Knut and Alice Wallenberg Foundation under Contracts Nos. 2021.0174, 2021.0299; Ministry of Development of Turkey under Contract No. DPT2006K-120470; National Research Foundation of Korea under Contract No. NRF-2022R1A2C1092335; National Science and Technology fund of Mongolia; Polish National Science Centre under Contract No. 2024/53/B/ST2/00975; Swedish Research Council under Contract No. 2019.04595; U. S. Department of Energy under Contract No. DE-FG02-05ER41374.

\bibliography{bibliography.bib}

\end{document}